\documentclass[sigconf]{acmart}

\AtBeginDocument{%
  }

\setcopyright{acmlicensed}
\copyrightyear{2025}
\acmYear{2025}
\acmDOI{XXXXXXX.XXXXXXX}
\acmConference[EASE 2025]{The 29th International Conference on Evaluation and Assessment in Software Engineering}{17–20 June, 2025}{Istanbul, Türkiye}
\acmISBN{978-1-4503-XXXX-X/18/06}

\begin{document}

\title{Critical Considerations on Effort-aware Software Defect Prediction Metrics}

\author{Luigi Lavazza}
\email{luigi.lavazza@uninsubria.it}
\orcid{0000-0002-5226-4337}
\affiliation{%
	\institution{Universit\`a degli Studi dell'Insubria}
	\city{Varese}
	\country{Italy}
}
\author{Gabriele Rotoloni}
\email{grotoloni@uninsubria.it}
\orcid{0000-0003-2046-0090}
\affiliation{%
	\institution{Universit\`a degli Studi dell'Insubria}
	\city{Varese}
	\country{Italy}
}
\author{Sandro Morasca}
\email{sandro.morasca@uninsubria.it}
\orcid{0000-0003-4598-7024}
\affiliation{%
	\institution{Universit\`a degli Studi dell'Insubria}
	\city{Varese}
	\country{Italy}
}

\renewcommand{\shortauthors}{Lavazza et al.}

\begin{abstract}
\emph{Background.} Effort-aware metrics (EAMs) are widely used to evaluate the effectiveness of software defect prediction models, while accounting for the effort needed to analyze the software modules that are estimated defective.
The usual underlying assumption is that this effort is proportional to the modules' size measured in LOC. However, the research on module analysis (including code understanding, inspection, testing, etc.) suggests that module analysis effort may be better correlated to code attributes other than size. \\
\emph{Aim.}
We investigate whether assuming that module analysis effort is proportional to other code metrics than LOC leads to different evaluations.\\
\emph{Method.}
We show mathematically that the choice of the code measure used as the module effort driver crucially influences the resulting evaluations.
To illustrate the practical consequences of this, we carried out a demonstrative empirical study, in which the same model was evaluated via EAMs, assuming that effort is proportional to either McCabe's complexity or LOC.\\
\emph{Results.}
The empirical study showed that EAMs depend on the underlying effort model, and can give quite different indications when effort is modeled differently. It is also apparent that the extent of these differences varies widely.\\
\emph{Conclusions.}
Researchers and practitioners should be aware that the reliability of the indications provided by EAMs depend on the nature of the underlying effort model. The EAMs used until now appear to be actually size-aware, rather than effort-aware: when analysis effort does not depend on size, these EAMs can be misleading.
\end{abstract}

\ccsdesc[500]{Mathematics of computing~Statistical graphics}
\ccsdesc[300]{Software and its engineering~Software defect analysis}

\keywords{Effort-aware Performance Metrics, Traditional Performance Metrics, Effort for software defectiveness analysis, Software Defect Prediction}


\maketitle

\section{Introduction}\label{sec:intro}
\begin{figure*}[h]
  \centering
  \includegraphics[scale=0.4]{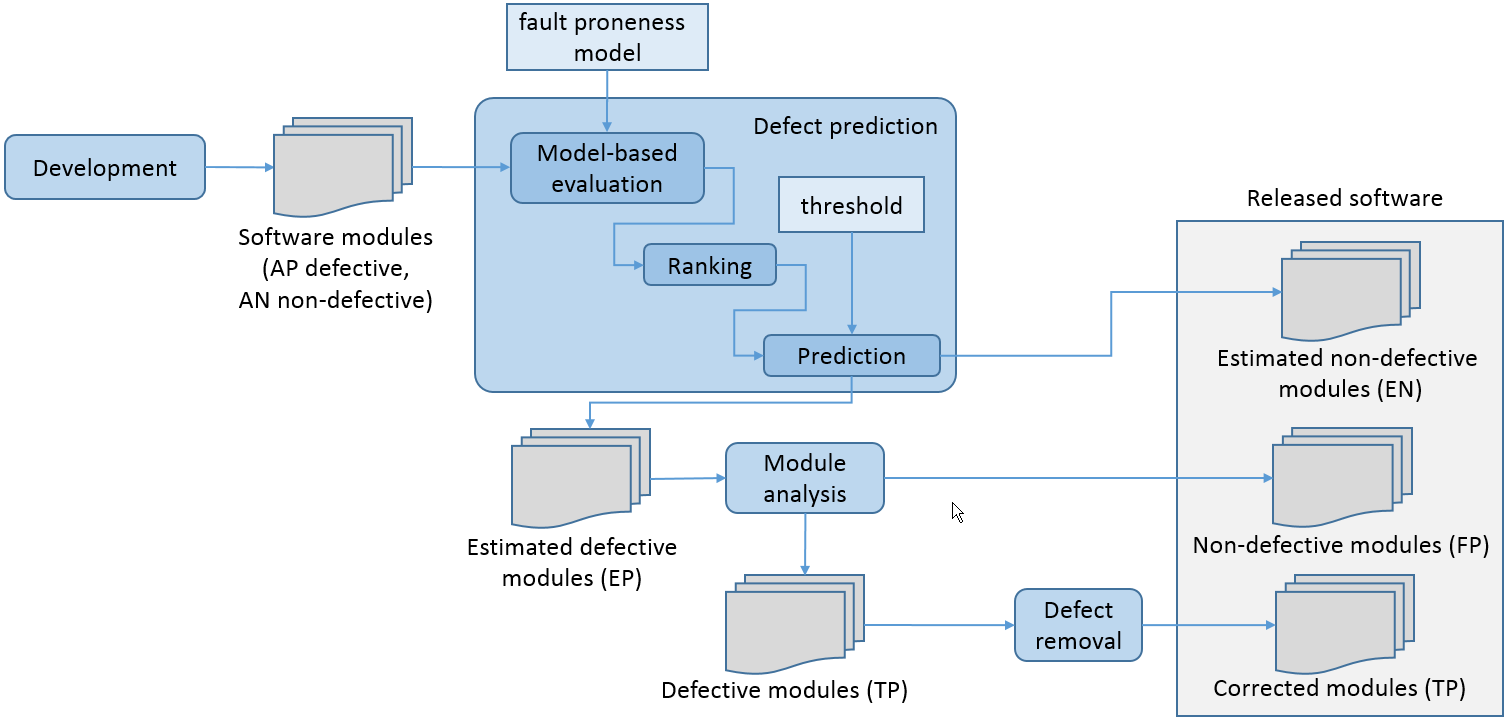}
  \caption{The process.}
  \label{fig:process}
  \Description{}
\end{figure*}

Software defect prediction (SDP) models aim to identify software modules\footnote{By ``module,'' we denote any piece of software (e.g., routine, method, class) whose defectiveness needs to be analyzed and evaluated.} that are likely to be defective, so defects can be located and removed before software is released to end-users.
Figure~\ref{fig:process} shows a typical quality assurance process that makes use of defect prediction.
Suppose a software system is made up of $n$ modules, $AP$ of which are actually positive (i.e., defective) and $AN$ actually negative (i.e., defect-free). Via the \texttt{Defect prediction} phase, $EN$ modules are estimated negative and are released without further analysis or modifications.
The other $EP\!=\!n\!-\!EN$ modules are estimated positive and undergo an additional quality assurance phase, \texttt{Module analysis}, to search for possible defects in them. As a result, out of the $EP$ estimated positives, $TP$ are found to be true positives, so their defects are removed and the modules are released. The other $FP\!=\!EP\!-\!TP$ modules are false positives, since the \texttt{Module analysis} phase did not find any defects in them and was thus unnecessarily carried out.

The effectiveness and efficiency of the process largely depend on the defect prediction phase, which relies on the use of SDP models, whose degree of correctness is thus of paramount importance.
SDP models are evaluated and selected via a number of performance
 metrics.
Traditional performance metrics (discussed in Section~\ref{sec:background}) do not consider the effort needed for \texttt{Module analysis}.
However, in industrial practice, the effort required to analyze estimated positive modules needs to be taken into account.
Notably, according to Mende and Koschke~\cite{mende2010effort}, when module analysis effort is considered, some SDP models may have a performance that appears to be reasonably good according to traditional metrics, while it is in fact worse than the performance obtained by picking the modules to analyze at random.
In addition, the effort required to analyze all the estimated positive modules may be beyond the amount of resources available, in practical situations.

Therefore, it is advisable to take module analysis effort into consideration when evaluating the performance of an SDP model.
Effort-aware metrics (EAM) were proposed for this purpose~\cite{rahman2012recalling,jiang2013personalized} and are gaining increasing attention.
The vast majority of the proposed EAMs assumes that analysis effort needed for a module is proportional to the size of the module, quantified in lines of code (LOC).

However, Mende and Koschke also suggested that module analysis can be correlated to other code attributes better than size~\cite{mende2010effort}.
An empirical study by Shihab et al. showed that LOC does not appear to be a reliable predictor of module analysis effort~\cite{shihab2013lines}: they found that complexity measures correlate with effort better than LOC, and concluded that effort-aware models should not assume that LOC is a good measure of effort.

In this paper, we investigate and evaluate the consequences of assuming that module analysis effort is proportional to other code measures than LOC.
Specifically, we mathematically show that a given EAM (like PofB~\cite{chen2017PofB}, for instance) is bound to give different indications when module analysis effort is considered proportional to different code measures.
In addition, we carried out an empirical study, to tangibly show the practical consequences of adopting different effort models.
To this end, we built SDP models and evaluated them via a few EAMs, assuming that effort is proportional to either size (measured in LOC) or structural code complexity (as measured by McCabe's cyclomatic complexity~\cite{McCabeTSE1976}).
Our paper shows that EAMs yield different values depending on the adopted effort model; that is, a given EAM will generally provide different performance evaluations for the same SDP model and the same analysis effort, when effort is assumed proportional to different code metrics.

The EAMs considered in this paper (more details in Section~\ref{sec:EAM}) are PofB~\cite{chen2017PofB}, Popt~\cite{mende2010effort}, and NPofB~\cite{ccarka2022effort}.

The paper is structured as follows.
Section~\ref{sec:background} describes the traditional approach to performance metrics, while Section~\ref{sec:EAM} describes  the foundations of the effort-aware approach to performance evaluation.
Section~\ref{sec:otherMetrics} discusses the consequences of assuming that effort depends on other qualities than size, and shows that EAMs that use different effort models are bound to provide different indications concerning performance.
Section~\ref{sec:consequences} illustrates the empirical study, showing the performance evaluations yielded by EAMs when effort is assumed proportional to McCabe's complexity instead of LOC. In this section, we also show the consequences of assuming that module analysis effort depends on \emph{both} LOC \emph{and} McCabe's complexity.
Section~\ref{sec:threats} discusses the threats to the validity of the study.
Section~\ref{sec:related} reports about the related work, while Section~\ref{sec:conclusion} draws some conclusions.

\section{Background}\label{sec:background}

For clarity, we use the term ``performance metric'' when quantifying the degree of correctness of a defectiveness model or defectiveness classifier. We use the term ``code measure'' or simply ``measure'' when referring to quantification of a quality of the software code being analyzed. A code measure is used to quantify an attribute of a software module~\cite{FentonBieman2014,BriandMorascaBasiliTSE1996}.
For instance, measure lines of code (LOC) is used to quantify attribute ``size,'' and measure McCabe's cyclomatic complexity (McCC) attribute ``complexity.'' Several different measures can be used to quantify an attribute. For instance, the size of a software class can be quantified by LOC, the number of statements, the number of methods, etc.

In this paper, we deal with the effort needed to analyze software modules. For conciseness' sake, we write ``effort'' instead of ``module analysis effort.''

The \texttt{Model-based evaluation} phase of the process in Figure~\ref{fig:process} estimates the likelihood of being faulty of each module. In the general case, this phase uses a scoring classifier~\cite{FawcettPRL2006}, i.e., a function that computes a score based on some code features, i.e., code measures. We call this scoring classifier the SDP model.
In Figure~\ref{fig:process} we introduced a \texttt{Ranking} phase to highlight that the scores provided by the SDP model can be ranked according to different criteria: for instance, if the SDP model provides an estimate of the modules' fault-proneness, one can simply order the modules by decreasing fault-proneness, but it is also possible to first divide the estimate by the module's size, before ordering, as is done when the NPofB EAM is used~\cite{ccarka2022effort} .
The estimation of whether a module is negative or positive takes place in the \texttt{Prediction} phase, which uses a \texttt{Threshold} in such a way that all modules with a score at least as high as the threshold are estimated positive and the others negative. Thus, setting a threshold on a defectiveness score results in a binary classifier, which we call a defectiveness classifier.

A special case of SDP model is a probabilistic classifier~\cite{FawcettPRL2006}, in which the score of a module is the estimated probability that the module is faulty. In this case, the threshold is the minimum probability required to estimate a module positive. A ``natural'' threshold is the proportion of actually positive modules, also known as prevalence and indicated by $\rho$ = AP/n, which is also the probability of selecting a defective module at random. The underlying idea is that a module should be considered defective if its probability of being defective is higher than the mean defectiveness probability of the set of modules, which is $\rho$. Several threshold setting criteria exist, based on the specific application at hand.
For instance, one could choose a rather low probability threshold to decrease the risk that defective modules are released.

In addition to the $EP$ estimated positives, split into $TP$ true positives and $FP$ false positives, the \texttt{Prediction} phase identifies $EN$ estimated negatives, which can be split into $TN$ (true negatives) correctly estimated actual negatives and the $FN$ (false negatives) incorrectly estimated actual positive modules. These modules are no longer dealt with in the software development process, but the release of false negative modules (which do contain faults) may lead to adverse consequences for the end-user.

$TP$, $FP$, $FN$, and $TN$ are the cells of the so-called confusion matrix, which has been used to define many performance metrics.  Table~\ref{tab:Performance_Metrics} reports some of the most commonly used performance metrics.

\renewcommand{\arraystretch}{1.5}
\begin{table}[hbt]
\centering
\caption{Some popular performance Metrics}
\label{tab:Performance_Metrics}
\begin{tabular}{ll}
\hline
Definition & Terms \\
\hline
$TPR=\frac{TP}{AP}$ & True Positive Rate, Recall, Sensitivity \\
\hline
$TNR=\frac{TN}{AN}$ &  True Negative Rate\\
\hline
$FPR=\frac{FP}{AN}$ & False Positive Rate , Specificity \\
\hline
$PPV=\frac{TP}{EP}$ & Positive Predicted Values , Precision \\
\hline
$Acc=\frac{TP+TN}{AP+AN}$ &  Accuracy \\
\hline
$BA=\frac{TPR+TNR}{2}$  &  Balanced Accuracy \\
\hline
$Gmean=\sqrt{TPR\cdot TNR}$ & Gmean\\
\hline
$F1=\frac{\textit{2}\ PPV\cdot TPR}{PPV + TPR}$ & F1, F-measure, F-score  \\
\hline
$MCC=\frac  {TP\cdot TN-FP \cdot FN}{{\sqrt  {EN \cdot EP \cdot AN \cdot AP}}}$ & Matthews Correlation Coefficient, $\phi$ \\
\hline
\end{tabular}
\end{table}
\renewcommand{\arraystretch}{1}


The confusion matrix and the corresponding performance metrics depend on the threshold chosen: different thresholds correspond to different defectiveness classifiers for the same defectiveness model. Therefore, performance metrics evaluate the specific defectiveness classifier corresponding to the threshold chosen, not the defectiveness model itself.

Performance metrics were also proposed to evaluate the entire defectiveness model.
The most common technique used for this purpose is based on the ROC (Receiver Operating Characteristic) curve, which depicts the performance of all the classifiers that can be obtained from a defectiveness model by varying the threshold: so, every point of the ROC curve is associated with a threshold and a confusion matrix.
Usually, the goodness of the ROC curve is summarized via its AUC (the area under the ROC curve).
Alternative metrics were also proposed: for instance, area under the recall-precision (TPR--PPV) curve is also widely used.

Most traditional performance metrics consider both the ability to detect defects and the amount of false alarms. The former are important to avoid releasing defective modules; the latter to avoid useless quality assurance activity.

\section{The Effort-aware Approach to Performance Evaluation}\label{sec:EAM}

Like the traditional techniques, effort-aware software defect prediction (EASDP) techniques require a defectiveness model.
PofB~\cite{chen2017PofB} is based on the module ranking used by the traditional approaches;
instead, NPofB (Normalized PofB)~\cite{ccarka2022effort} ranks modules according to their defectiveness density, obtained dividing the defectiveness score of a module by its size.

Given the ranking, the threshold is set based on the maximum effort available for analysis. Thus, all the highest ranking modules whose cumulative size requires no more than the available effort are estimated positive and analyzed.

Without any loss of generality, we can always suppose that the modules are ordered in monotonically non-increasing rank order according to a defectiveness model, i.e., the rank of the $i^{th}$ module is never lower than the rank of the $(i+1)^{th}$ module.
Each module has a set of attributes, including size: we denote by $s_i$ the size in LOC of the $i^{th}$ module.

EAMs use LOC as a proxy of the analysis effort, i.e., it is assumed that the effort to analyze the $i^{th}$ module is proportional to its size measured in LOC $s_i$
\begin{equation}\label{eq:effortS}
\emph{effort}_i(LOC)=c_{LOC}\cdot s_i
\end{equation}
$c_{LOC}$ is the cost of analyzing a single LOC, expressed in PersonHours/LOC or some similar unit.
We write $\emph{effort}_i(LOC)$ instead of just $\emph{effort}_i$ to stress that this effort is assumed to depend on size, measured in LOC.

The effort needed to analyze the entire system is
\[
\emph{effort}_{\emph{Sys}}(LOC)=\sum_{i=1}^n c_{LOC}\cdot s_i=c_{LOC}\cdot \sum_{i=1}^n s_i
\]
since $\sum_{i=1}^n s_i$ is the size of the entire system.

The maximum allowable effort for phase \texttt{Module analysis} in Figure~\ref{fig:process} is used to determine which modules are estimated positive. Suppose that up to $EP$ highest-ranking modules can undergo module analysis with the available effort. Their cumulative analysis effort is $\sum_{i=1}^{EP} c_{LOC}\cdot s_i$, which could be lower than the available effort. However, adding an extra module, i.e., analyzing $(EP\!+\!1)$ modules, would exceed the available effort, since \texttt{Module analysis} applies only to entire modules. So, the available resources effectively allow a fraction $t_{LOC}$ of the effort needed for the entire system to be employed
\begin{equation}\label{eq:effortThreshold}
t_{LOC}=\frac{\sum_{i=1}^{EP} c_{LOC}\cdot s_i}{\sum_{i=1}^n c_{LOC}\cdot s_i}=\frac{\sum_{i=1}^{EP} s_i}{\sum_{i=1}^n s_i}
\end{equation}

Equation (\ref{eq:effortThreshold}) shows that $t_{LOC}$ can be considered a fraction of the effort needed to analyze the entire system or of the total size of the system. Thus, the amount of available effort induces a threshold $EP$ on the number of analyzable modules according to the defectiveness model used, which, in turn, induces a threshold $t_{LOC}$ on the fraction of effort used or, equivalently, of analyzed LOC. Conversely, threshold $t_{LOC}$ induces a threshold $EP$ on the number of analyzable modules according to the defectiveness model used.
In this paper, we regard $t_{LOC}$ as a fraction of effort and we write ``\% effort(LOC)'' instead of $t_{LOC}$, to stress that we are dealing with an effort ratio.

Of the $EP$ modules that are analyzed, $TP$ are actually defective and $FP$ are actually non-defective.
Likewise, of the remaining $EN$ modules, $TN$ are non-defective and $FN$ are defective.
Therefore, a confusion matrix is obtained.

Most EAMs, including PofB, NPofB and Popt, indicate the proportion of defective modules detected (the \texttt{Defective modules} in Figure~\ref{fig:process}), i.e., TPR of Table~\ref{tab:Performance_Metrics}.
Some EAMs---although adopting the process depicted in Figure~\ref{fig:process} and assuming effort proportional to LOC---
use the resulting confusion matrix to compute different values, like the F-measure~\cite{cheng2022effort} or PPV~\cite{li2023impact}

Like with traditional performance metrics, an EAM value depends on the threshold: for instance, the value of PofB with $t_{LOC} = 0.5$, known as PofB50 in the literature, is never less and generally greater than PofB20, the value of PofB with $t_{LOC} = 0.2$, since the set of modules estimated positive when $t_{LOC} = 0.5$ includes the set of modules estimated positive when $t_{LOC} = 0.2$.

Like in the traditional case, in EASDP some indicators that do not depend on a specific value of the threshold are also available. The most commonly used is Popt~\cite{mende2010effort}. Like the AUC, Popt considers an area under a curve, but the curve is a cost efficiency curve, which is a plot of the performance (expressed via TPR) as a function of the effort dedicated to module analysis, as we show in Section~\ref{subsec:PoptComparisons}.

\section{Assuming that Effort Depends on Other Qualities than Size}\label{sec:otherMetrics}

EAMs aim to consider the cost of module analysis (\texttt{Module analysis} in Figure~\ref{fig:process}), but they do not consider how module analysis is performed.
Since module analysis can include a variety of activities (static code analysis, code inspection, testing, etc.), its cost may depend on multiple factors. For instance, the cost of code inspection depends largely on the size of the code to be analyzed~\cite{arisholm2006predicting,arisholm2007data}, while the cost of testing depends mostly on the code structure as measured by McCabe's complexity~\cite{bruntink2004predicting,bruntink2006empirical}.
More recent research on code understandability has also shown that other factors are as relevant as size~\cite{munoz2020empirical,lavazza2023empirical}. 
Since code understanding is part of analysis activities~\cite{oliveira2024understanding}, it is reasonable to doubt whether size is the only, or even the main, driver of effort.
Therefore, assuming that analysis effort is proportional to size appears as a limitation of EAMs.

In this section, we consider different hypotheses concerning the attributes of code that determine analysis effort.
Accordingly, we investigate the use of measures other than LOC as drivers for module analysis and its consequences.

Note that we do not make any assumption concerning the used SDP model or the criterion to rank modules.
Therefore, the following observations apply to any combination of SDP model and ranking method.

\subsection{Generalizing EAMs with Respect to Effort Drivers}\label{subsec:generalization}

We redefine the effort required to analyze the $i^{th}$ module as
\begin{equation}\label{eq:effort_q}
\emph{effort}_i(M)=c_M\cdot m_i
\end{equation}
where $M$ is a measure for some code attribute that is known as an effort driver, $m_i$ is the value of measure $M$ for the $i^{\emph{th}}$ module, and $c_M$ is the effort needed to analyze a piece of code with one unit of code measure $m$. This is a straightforward generalization of the size-based effort-aware approach of Section \ref{sec:EAM} to other attributes and measures.

Threshold $t_M$ used to classify modules using $M$ is defined as
\begin{equation}\label{eq:effort_threshold_q}
t_M=\frac{\sum_{i=1}^{EP} c_M\cdot m_i}{\sum_{i=1}^n c_M\cdot m_i}=\frac{\sum_{i=1}^{EP} m_i}{\sum_{i=1}^n m_i}
\end{equation}

Equations (\ref{eq:effortS}) and (\ref{eq:effortThreshold}) are special cases of (\ref{eq:effort_q}) and (\ref{eq:effort_threshold_q}), when $M$ is size measured in LOC.
In fact, we do not change the nature of EAMs, instead we let effort be determined by any code measure.

For clarity, we express $t_M$ as a percentage in what follows, and we name it ``\%effort(M):'' for instance, \%effort(McCC) indicates a fraction of the effort needed to analyze the entire system, when the effort is assumed proportional to McCabe's complexity (McCC).

\subsection{Properties of EAMs}\label{subsec:properties}

Let us consider an EAM measured on two effort models: one that assumes that the effort required to analyze the $i^{th}$ module is $\emph{effort}_i\!=\!c_{LOC}\!\cdot\!s_i$ and another that assumes $\emph{effort}_i\!=\!c_Q\cdot\!q_i$, where $Q$ is a code measure, like McCabe's cyclomatic complexity (McCC), that is expected to determine the analysis effort.
For the EAM to be valid in both scenarios, we need both effort models to be correct.
This implies, among other things, that both models represent correctly the effort needed to analyze the entire system:
\[
\emph{effort}_{Sys}=\sum_{i=1}^n c_{LOC}\cdot s_i =  \sum_{i=1}^n c_Q\cdot q_i
\]
hence
\begin{equation}\label{eq:ratio}
\frac{c_{LOC}}{c_Q}=\frac{\sum_{i=1}^n q_i}{\sum_{i=1}^n s_i}
\end{equation}

Now, suppose that the effort available for module analysis is $\overline{\emph{effort}} \in [0,\emph{effort}_{Sys}]$.
The available effort determines the number of modules EP that are analyzed:
\begin{equation}\label{eq:equalEffort}
\overline{\emph{effort}}=\sum_{i=1}^{EP_{LOC}} c_{LOC}\cdot s_i =  \sum_{i=1}^{EP_Q} c_Q\cdot q_i
\end{equation}

Now, two possibilities exist: either $EP_{LOC}\!=\!EP_Q$ or $EP_{LOC}\!\neq\!EP_Q$.

Let us consider the case when $EP_{LOC}=EP_Q=EP$: it is
\begin{equation}\label{eq:ESEPQP}
\overline{\emph{effort}}=\sum_{i=1}^{EP} c_{LOC}\cdot s_i =  \sum_{i=1}^{EP} c_Q\cdot q_i
\end{equation}

Since $\overline{\emph{effort}}$ can assume any value in $[0,\emph{effort}_{Sys}]$, EP can assume any value in $[1,n]$ (not considering the trivial case when the available effort is not event sufficient to analyze one module).
Therefore, it must be

\begin{eqnarray}
\forall\!j\!\in\![1,n]\  \sum_{i=1}^j c_{LOC}\cdot s_i =  \sum_{i=1}^j c_Q\cdot q_i \Leftrightarrow \nonumber \\
\forall\!j\!\in [1,n]\ c_{LOC}\!\sum_{i=1}^j s_i = c_Q\!\sum_{i=1}^j  q_i
\end{eqnarray}

Hence,
\begin{equation}\label{eq:equal_metrics_cumul}
\forall\!j\!\in\![1,n]\  c_{LOC}\cdot s_j =  c_Q\cdot q_j \Leftrightarrow \forall\!j\!\in\![1,n]\ s_j = \frac{c_Q}{c_{LOC}} q_j
\end{equation}

Because of Equation (\ref{eq:ratio}), $\frac{c_Q}{c_{LOC}}$ is constant, hence Equation (\ref{eq:equal_metrics_cumul}) is satisfied only if the measures of LOC and Q are always proportional, i.e., if they are the same measure from a practical standpoint, like LOC and KLOC (Kilo LOC).

Let us now consider the cases when $EP_{LOC}\!\neq\!EP_Q$ in Equation (\ref{eq:equalEffort}).
This is a situation that could not occur, if both effort models were correct.
In fact, in the real world, the available effort $\overline{\emph{effort}}$ is spent to analyze $\overline{EP}$ modules: hence, correct effort models would result in $EP_{LOC}\!=\!EP_Q\!=\!\overline{EP}$. If $EP_{LOC}\!\neq\!EP_Q$, at least one model is not correct.

In conclusion, either equation (\ref{eq:ESEPQP}) holds for all values of $\overline{\emph{effort}}$, hence Q and LOC are essentially the same measure, or, within the many possible values of $\overline{\emph{effort}}$, some are not correctly modeled via LOC or Q (or both).

Intuitively, the latter situation results in different proportions of defective modules reported by EAMs using different effort models, since a greater EP means that more modules are analyzed, hence more defects are found, in general.
The empirical study described in the next Section~\ref{sec:consequences} shows that this is precisely what happens when McCabe's complexity is used as Q.

\section{Practical Consequences}\label{sec:consequences}

This section illustrates the practical consequences of assuming that effort is proportional to other code attributes than size.


Among the many qualities of code, complexity---as measured by McCabe's cyclomatic complexity (McCC)~\cite{McCabeTSE1976}---is one of the most important effort drivers.
It is has also been reported that testing effort appears to be correlated with McCC
~\cite{bruntink2004predicting,bruntink2006empirical}, hence considering McCC as an effort driver for module analysis seems reasonable.
It is also well known that McCC is quite often well correlated to size expressed in LOC.
Therefore, we consider cyclomatic complexity as a possible alternative effort driver to be used as $Q$.

\subsection{The Empirical Study}\label{subsec:study}
In this section, we illustrate via examples the consequences of assuming that analysis effort is proportional to code measures different from LOC.
Note that this section is purely demonstrative. The phenomena illustrated here descend from the properties of EAMs described in Section~\ref{sec:otherMetrics}.

\subsubsection{The Dataset}\label{subsubsec:dataset}
To carry out the analysis, we used 12 datasets from the NASA collection~\cite{shepperd2013data}, which was also used by Mende and Koschke~\cite{mende2010effort} in their seminal work.
The cleaned NASA datasets are available from GitHub (https://github.com/klainfo/ NASADefectDataset); each dataset contains the value of several (between 20 and 37) code measures for each module, along with the indication of whether the module contains any defect or not.

\subsubsection{Model Building}\label{subsubsec:model}

With reference to Figure~\ref{fig:process}, the computation of EAMs involves \texttt{Ranking} and \texttt{Prediction},
using whatever SDP model is yielded by \texttt{Model-based evaluation}.
Therefore, for our purposes, the technique used to obtain the prediction models is inessential.
In this study, we used Binary Logistic Regression (BLR) to build models.
We chose BLR because it yields proper defectiveness probability models that are explainable.

%
%
%

All the models used in the following sections used size (LOC) and McCabe complexity density (i.e., McCC/LOC) as independent variables. However, in Section~\ref{subsubsec:differentModel} a different model is used, to show that the presented results do not depend on a specific model.

\subsubsection{Evaluation}\label{subsubsec:eval}

To show the differences among the values that EAMs yield when different effort drivers are adopted, we use cost efficiency curves similar to those used by Mende and Koschke~\cite{mende2010effort}.
However, since our models provide defectiveness probability models, while Mende and Koschke used models yielding the estimated number of defects per module, our cost efficiency curves describe the proportion of defective modules detected
rather than the proportion of defects found.
An example of these curves can be found in Figure~\ref{fig:PC3_LOC_TOTAL+CYCLOMATIC_DENSITY}.
Cost efficiency curves have the percentage of effort allocated to module analysis on the $x$ axis, and TPR (the fraction of defective modules found via module analysis) on the $y$ axis.

Because of space limitations, we report cost efficiency curves for just two datasets. The other curves---which show similar results---are available in the supplemental material.

\subsection{Effort Computation Affects the Performance Metrics}\label{subsec:res_effort}

\begin{figure}[h]
  \centering
  \includegraphics[scale=0.55]{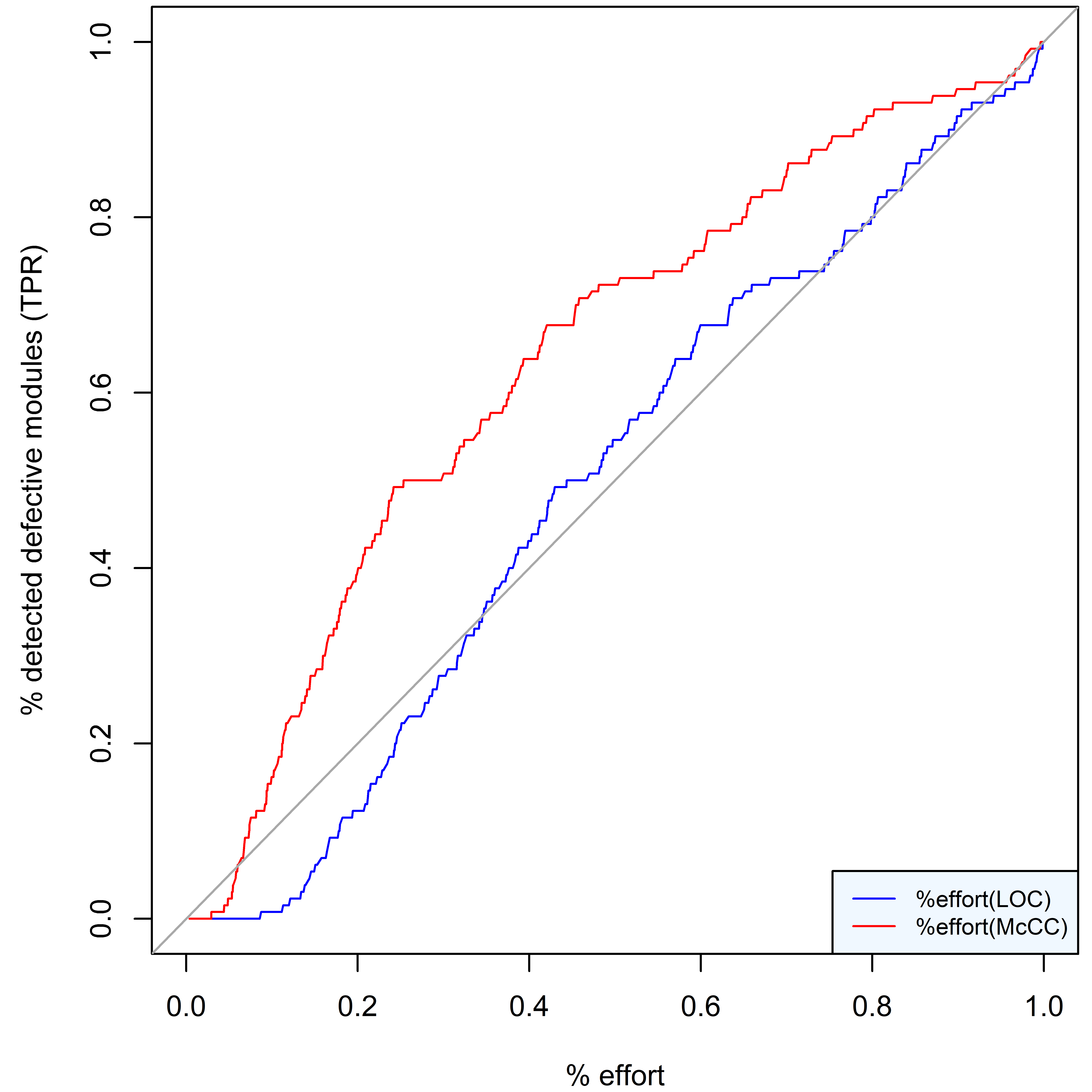}
  \caption{Cost efficiency curves of the BLR SDP model that uses LOC and McCabe's complexity density as independent variables for project \texttt{PC3}.}
  \label{fig:PC3_LOC_TOTAL+CYCLOMATIC_DENSITY}
  \Description{}
\end{figure}

Figure~\ref{fig:PC3_LOC_TOTAL+CYCLOMATIC_DENSITY} shows the cost efficiency curves of a model, when analysis effort is assumed to be proportional to the size in LOC or to McCabe's complexity.

Figure~\ref{fig:PC3_LOC_TOTAL+CYCLOMATIC_DENSITY} shows that assuming that analysis effort is proportional to McCabe's complexity can lead to very different evaluations with respect to assuming that analysis effort is proportional to the size expressed in LOC.
If effort is proportional to LOC, allocating 20\% of the effort needed to analyze the entire system appears to perform worse than picking modules at random (the blue line is below the diagonal when \%effort(LOC)=20\%).
Instead, if effort is proportional to McCabe complexity, \%effort(McCC)=20\% allows the detection of around 40\% of the defective modules.

In the empirical study, we used only LOC and McCC as effort drivers because i) they are the only metrics that were used for this purpose in the literature~\cite{mende2010effort},
 ii) we could use only measures from the NASA dataset that are expected to affect analysis effort, and iii) LOC and McCC are sufficient to show that quite different performance evaluations can be obtained via EAMs.

\subsection{PofB Comparisons}\label{subsec:PofBComparisons}

Table~\ref{tab:pofb2050} shows the values of PofB20 and PofB50 obtained for all projects, when either LOC or McCC is used as cost drivers.

\renewcommand{\arraystretch}{1}
\begin{table}[h]
\centering
\caption{Comparison of PofB20 and PofB50 when LOC and McCC are used as effort drivers.}
\label{tab:pofb2050}
\begin{tabular}{l|cc|cc}
\hline
        & \multicolumn{2}{c}{PofB20}    & \multicolumn{2}{c}{PofB50}    \\
Project & LOC           & McCC          & LOC           & McCC          \\
\hline
\texttt{CM1}     & 0.07          & 0.12          & 0.48          & 0.62 \\
\texttt{JM1}     & 0.07          & 0.08          & 0.30          & 0.36 \\
\texttt{KC1}     & 0.09          & 0.09          & 0.31          & 0.34 \\
\texttt{KC3}     & 0.11          & 0.17          & 0.36          & 0.47 \\
\texttt{MC1}     & 0.11          & 0.11          & 0.47          & 0.58 \\
\texttt{MC2}     & 0.09          & 0.09          & 0.27          & 0.27 \\
\texttt{PC1}     & 0.24          & 0.33          & 0.58          & 0.78 \\
\texttt{PC2}     & 0.19          & 0.44          & 0.63          & 0.75 \\
\texttt{PC3}     & 0.12          & 0.39          & 0.55          & 0.72 \\
\texttt{PC4}     & 0.15          & 0.35          & 0.53          & 0.80 \\
\texttt{PC5}     & 0.02          & 0.05          & 0.09          & 0.38 \\
\hline
\end{tabular}
\end{table}
\renewcommand{\arraystretch}{1}

\begin{figure}[h]
  \centering
  \includegraphics[scale=0.55]{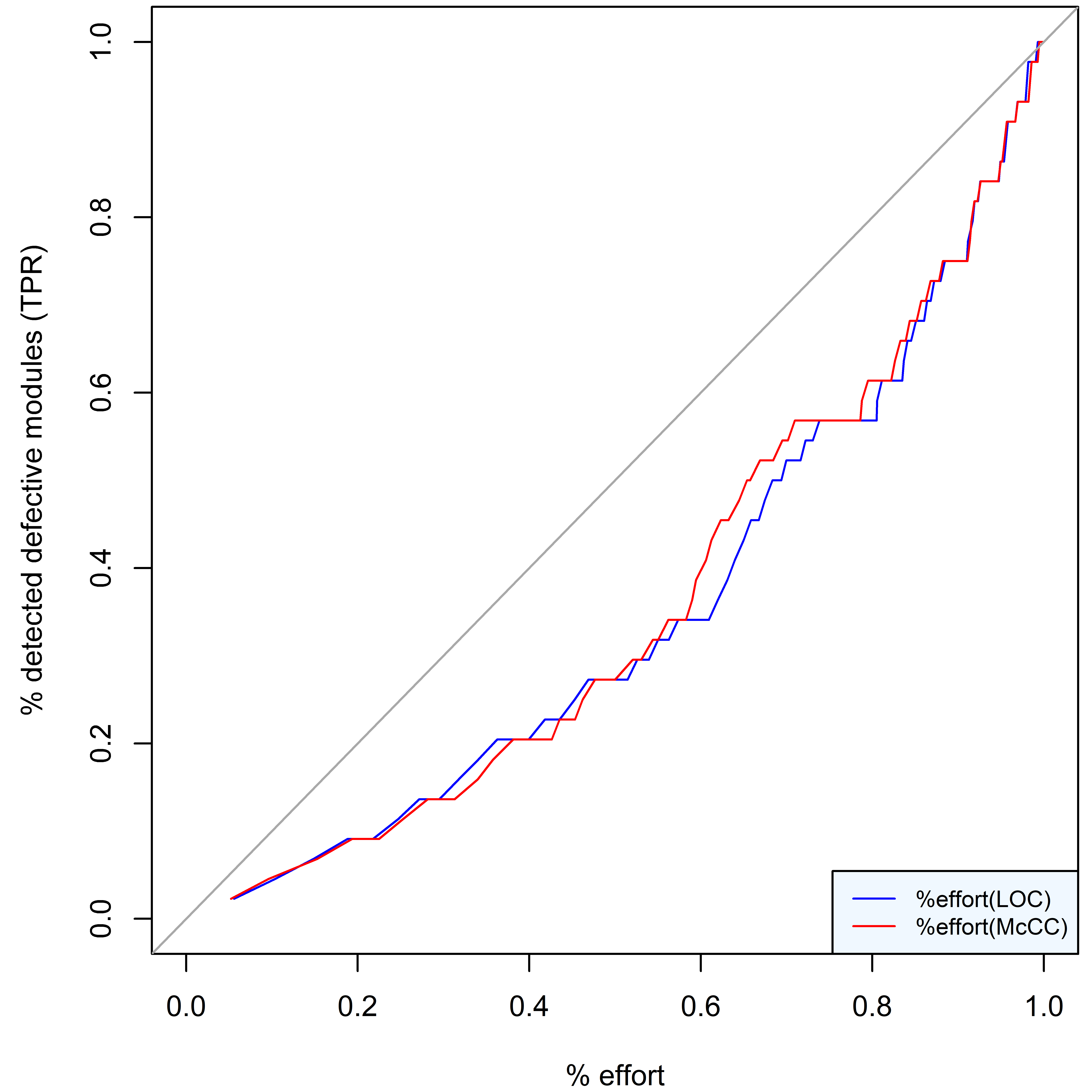}
  \caption{Cost efficiency curves of the BLR SDP model for project \texttt{MC2}, when LOC and McCabe's complexity density are used as independent variables.}
  \label{fig:MC2_LOC_TOTAL+CYCLOMATIC_DENSITY}
  \Description{}
\end{figure}

It can be seen that in general PofBX values are different, when different cost drivers are used.
The only exception is project \texttt{MC2}, whose cost efficiency curves, shown in Figure~\ref{fig:MC2_LOC_TOTAL+CYCLOMATIC_DENSITY}, are very close to each other, even though clearly non coincident.
They also cross each other multiple times.

The differences shown in Table~\ref{tab:pofb2050} were expected, as they are a direct consequence of Section~\ref{sec:otherMetrics}.

\subsection{Optimal Cost Efficiency Curves}\label{subsec:OptimalCurves}

Optimal cost efficiency curves are built by considering the defective modules first, in order of growing size (or any other effort driver), and then the non-defective ones.

\begin{figure}[h]
	\centering
	\includegraphics[scale=0.55]{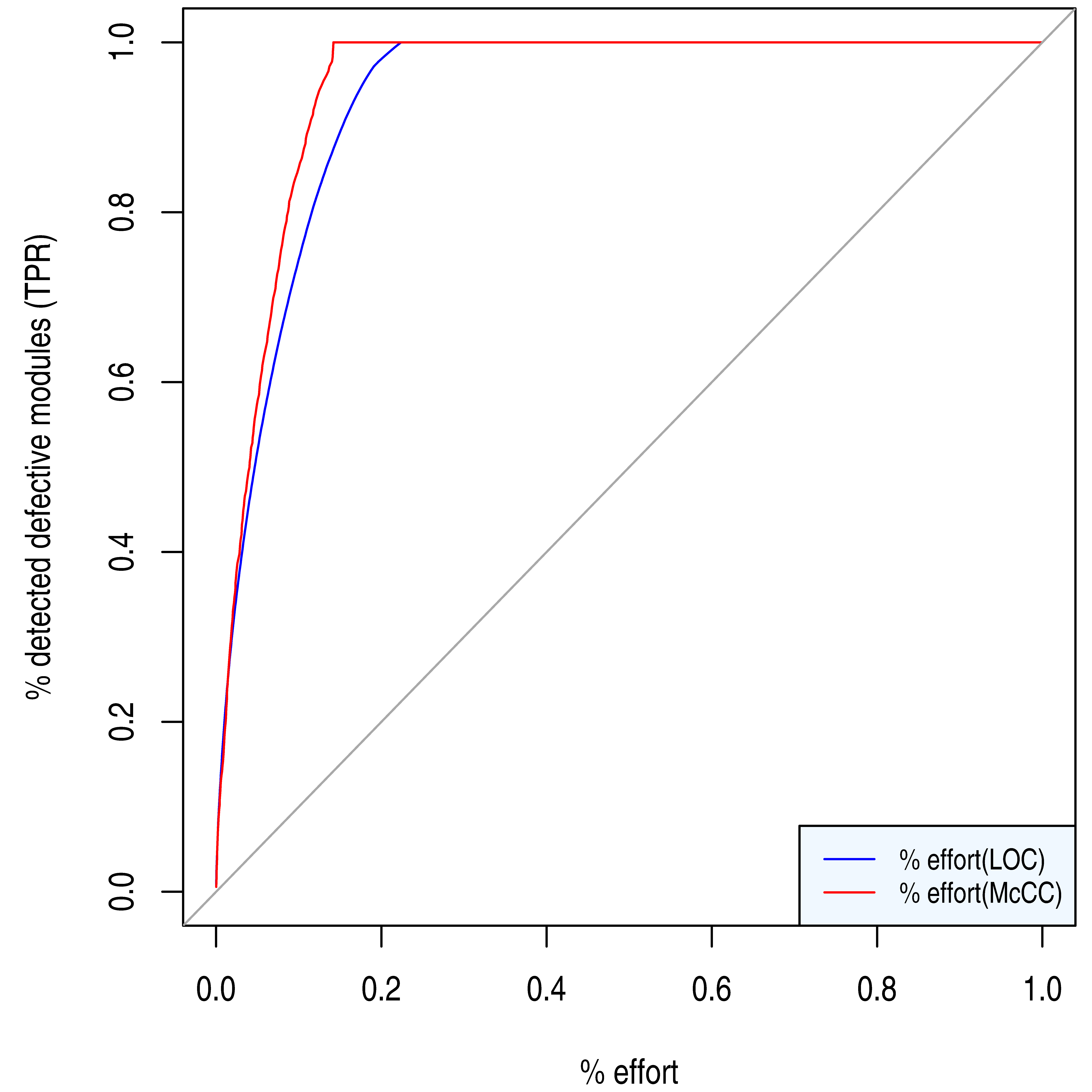}
	\caption{Optimal cost efficiency curves for project \texttt{PC4}, when LOC and McCabe's complexity are used as effort drivers.}
	\label{fig:PC4_ideal}
	\Description{}
\end{figure}

\begin{figure}[h]
	\centering
	\includegraphics[scale=0.55]{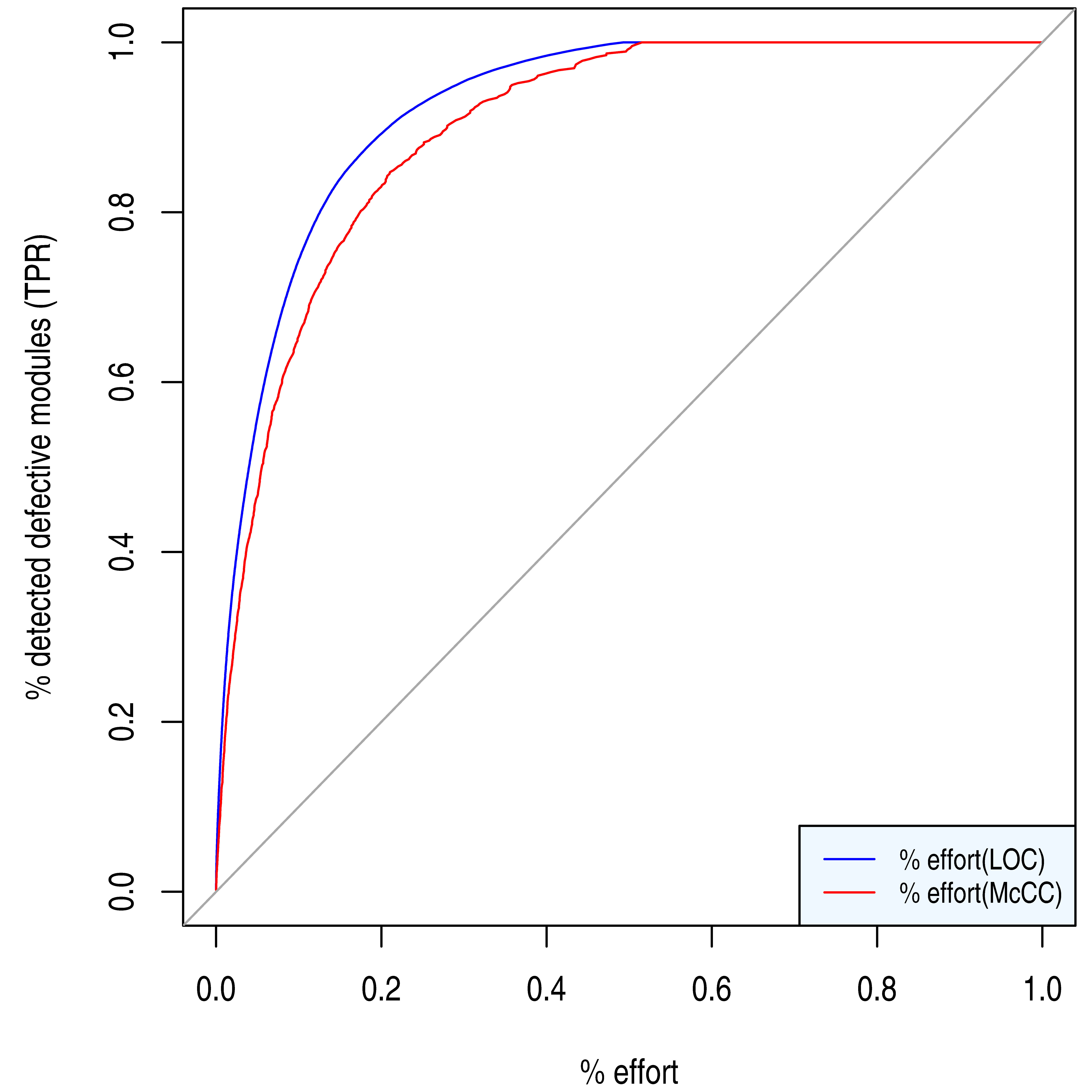}
	\caption{Optimal cost efficiency curves for project \texttt{PC5}, when LOC and McCabe's complexity are used as effort drivers.}
	\label{fig:PC5_ideal}
	\Description{}
\end{figure}

Figures~\ref{fig:PC4_ideal} and~\ref{fig:PC5_ideal} show the optimal efficiency curves for projects \texttt{PC4} and \texttt{PC5}, respectively, when either size (LOC) or McCabe complexity are used as effort drivers.
The best performance is obtained when analysis effort is assumed to be proportional to LOC for \texttt{PC5}, when the effort is assumed to be proportional to McCC for \texttt{PC4}.

This fact has an important implication. In fact, in principle one could wonder if the results described in this paper depend on the defectiveness models; i.e., ``optimized'' models could support consistent EAM values with different effort drivers.
Figures~\ref{fig:PC4_ideal} and~\ref{fig:PC5_ideal} show that it is not so: even the best possible models, i.e., those providing rankings that are equivalent to the optimal ones, yield different EAM values with different effort drivers.


\subsection{Popt Comparisons}\label{subsec:PoptComparisons}
Popt was defined by Mende and Koschke as $1\!-\!\Delta_{opt}$, where $\Delta_{opt}$ is the area between the optimal model (see Section~\ref{subsec:OptimalCurves}) and the considered model.
So, the closer the considered model to the optimal model, the lower $\Delta_{opt}$ and the higher Popt.

The definition of Popt aims to provide a model evaluation that does not depend on a specific threshold, as Popt considers all the possible amounts of effort that can be dedicated to module analysis, from nil to the amount needed to analyze the entire system.

We computed Popt for the same models used in the previous sections, for all of the NASA projects.
The results are in Table~\ref{tab:comp_popt}.

\renewcommand{\arraystretch}{1}
\begin{table}[h]
	\centering
	\caption{Comparison of Popt values when LOC and McCC are used as effort drivers.}
	\label{tab:comp_popt}
		\begin{tabular}{l|cc}
			\hline
			& \multicolumn{2}{c}{Popt} \\
			Project & LOC & McCC  \\
			\hline
			\texttt{CM1}     & 0.54     & 0.60  \\
			\texttt{JM1}     & 0.44     & 0.48  \\
			\texttt{KC1}     & 0.47     & 0.49  \\
			\texttt{KC3}     & 0.48     & 0.54  \\
			\texttt{MC1}     & 0.43     & 0.52  \\
			\texttt{MC2}     & 0.47     & 0.49  \\
			\texttt{PC1}     & 0.63     & 0.71  \\
			\texttt{PC2}     & 0.56     & 0.69  \\
			\texttt{PC3}     & 0.55     & 0.68  \\
			\texttt{PC4}     & 0.59     & 0.72  \\
			\texttt{PC5}     & 0.32     & 0.50  \\
			\hline
		\end{tabular}
\end{table}
\renewcommand{\arraystretch}{1}

The Popt values obtained with McCC as the cost driver are uniformly better than those obtained with LOC.
The difference varies from quite small (e.g., for projects \texttt{KC1} and \texttt{MC2}) to quite relevant (e.g., for projects \texttt{PC3}, \texttt{PC4} and \texttt{PC5}).


\subsection{Effort Computation Affects NPofB}\label{subsec:npofb}
In this section, we address NPofB~\cite{ccarka2022effort}, which differs from PofB only in the \texttt{Ranking} phase (see Figure~\ref{fig:process}), as it ranks modules according to their ``normalized'' defectiveness probability, i.e., fault proneness divided by LOC.

The results of this section were obtained using the same defectiveness models used in Sections~\ref{subsec:res_effort} and~\ref{subsec:PofBComparisons}.

\begin{figure}[h]
  \centering
  \includegraphics[scale=0.55]{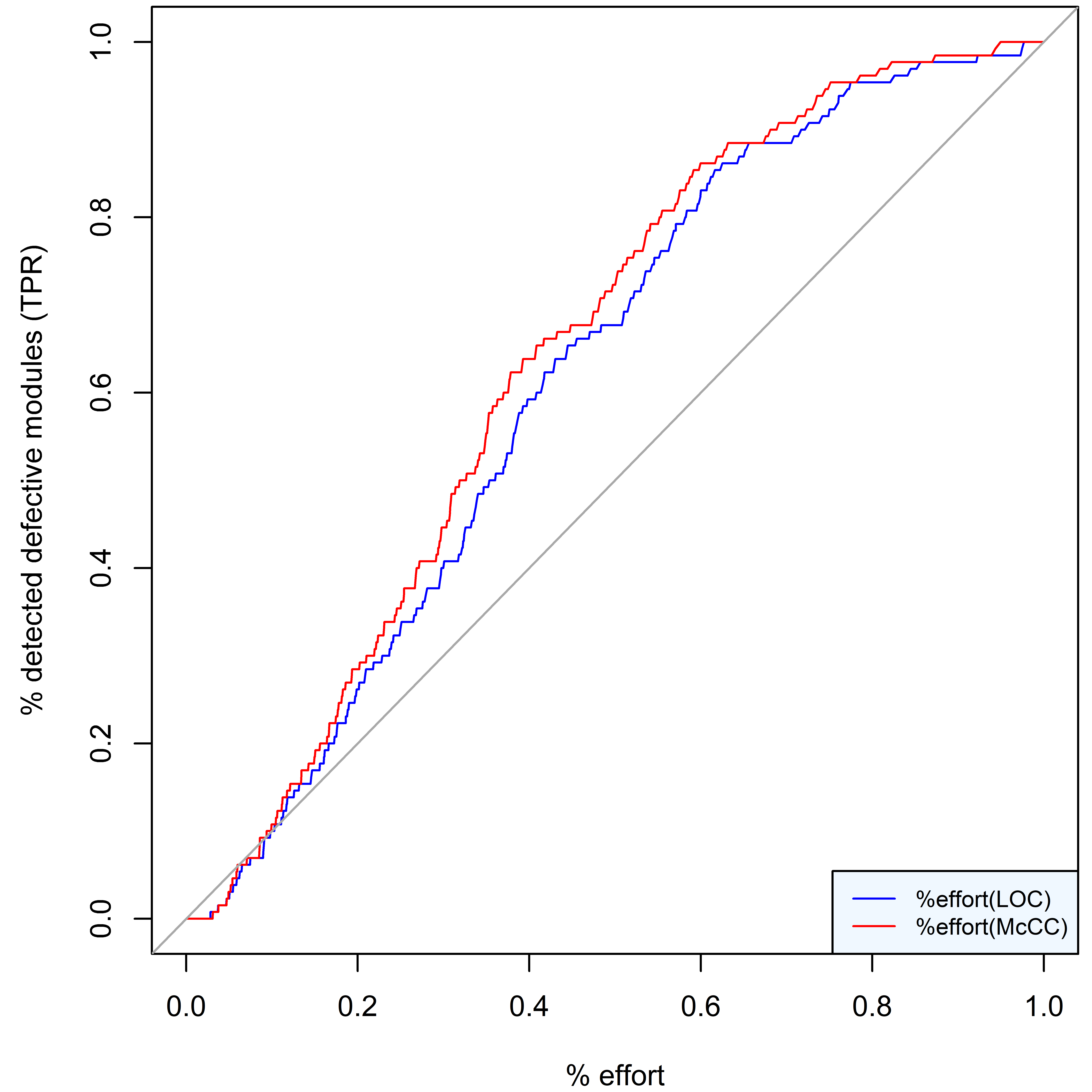}
  \caption{NPofB cost efficiency curves of the BLR SDP model that uses LOC and McCabe's complexity density as independent variables for project \texttt{PC3}.}
  \label{fig:PC3_NORM_LOC_LOC_TOTAL+CYCLOMATIC_DENSITY}
  \Description{}
\end{figure}

\begin{figure}[h]
  \centering
  \includegraphics[scale=0.55]{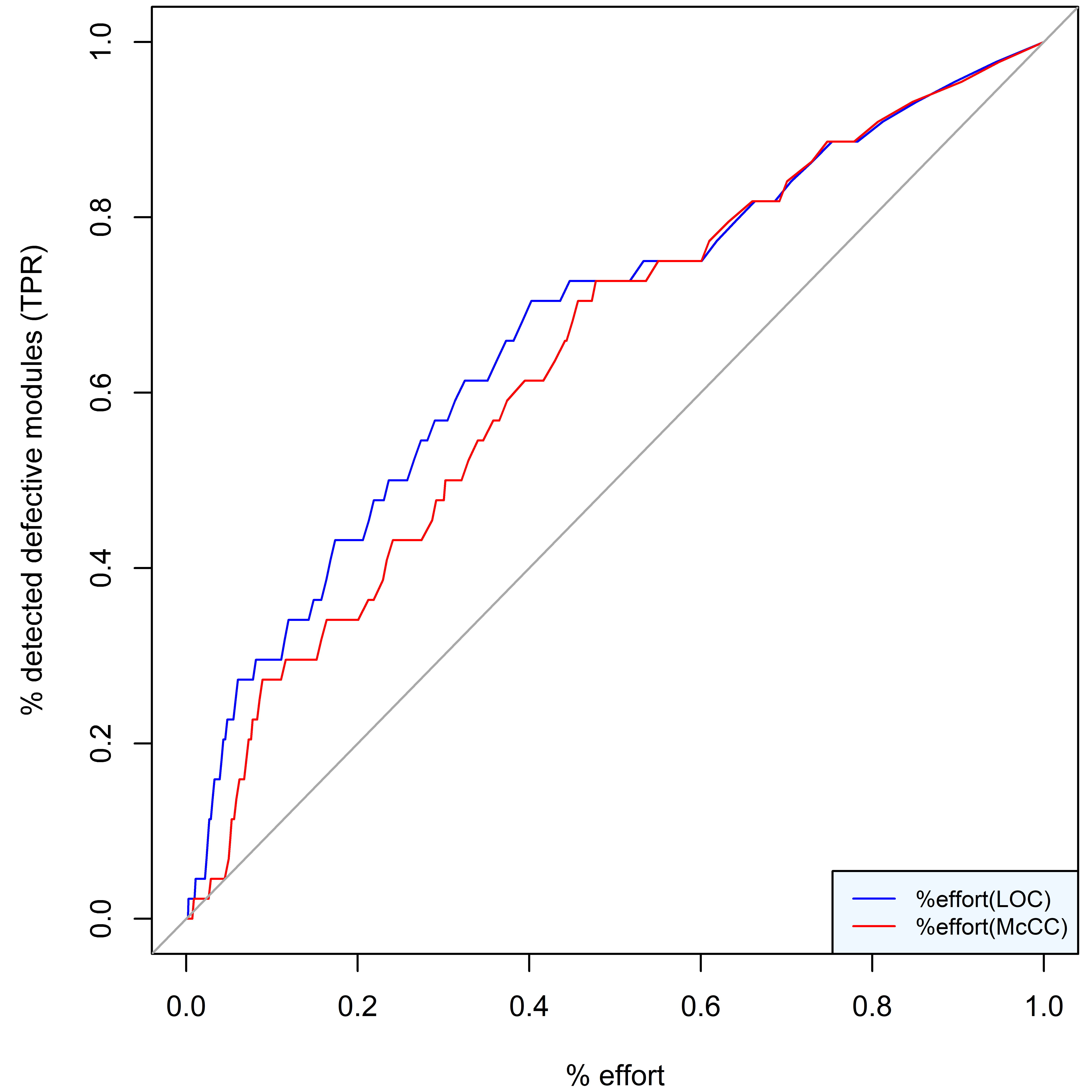}
  \caption{NPofB cost efficiency curves of the BLR SDP model that uses LOC and McCabe's complexity density as independent variables for project \texttt{MC2}.}
  \label{fig:MC2_NORM_LOC_LOC_TOTAL+CYCLOMATIC_DENSITY}
  \Description{}
\end{figure}

Figures~\ref{fig:PC3_NORM_LOC_LOC_TOTAL+CYCLOMATIC_DENSITY} and~\ref{fig:MC2_NORM_LOC_LOC_TOTAL+CYCLOMATIC_DENSITY} illustrate cost efficiency curves when NPofB is used.
Even though the curves are closer to each other than the corresponding PofB-based curves in   Figures~\ref{fig:PC3_LOC_TOTAL+CYCLOMATIC_DENSITY} and~\ref{fig:MC2_LOC_TOTAL+CYCLOMATIC_DENSITY}, it is apparent that we get different indications when effort is considered proportional to different code measures.

Table~\ref{tab:Npofb2050} provides the NPofB20 and NPofB50 values for all projects, when LOC and McCC are used as effort drivers.
Like in Table~\ref{tab:pofb2050}, NPofBX values are generally different when different effort drivers are used. Specifically, differences range from nil (project MC1) to definitely relevant (project \texttt{PC1}).

\begin{table}[]
\centering
\caption{Comparison of NPofB20 and NPofb50 when LOC and McCC are used as effort drivers.}
\label{tab:Npofb2050}
\begin{tabular}{l|cc|cc}
\hline
        & \multicolumn{2}{c}{NPofB20} & \multicolumn{2}{c}{NPofB50} \\
Project & LOC          & McCC        & LOC          & McCC        \\
\hline
\texttt{CM1}     & 0.38         & 0.40        & 0.71         & 0.79        \\
\texttt{JM1}     & 0.35         & 0.31        & 0.69         & 0.67        \\
\texttt{KC1}     & 0.35         & 0.33        & 0.70         & 0.71        \\
\texttt{KC3}     & 0.39         & 0.33        & 0.61         & 0.58        \\
\texttt{MC1}     & 0.25         & 0.25        & 0.58         & 0.58        \\
\texttt{MC2}     & 0.43         & 0.34        & 0.73         & 0.73        \\
\texttt{PC1}     & 0.33         & 0.51        & 0.71         & 0.87        \\
\texttt{PC2}     & 0.31         & 0.38        & 0.69         & 0.69        \\
\texttt{PC3}     & 0.26         & 0.28        & 0.68         & 0.72        \\
\texttt{PC4}     & 0.36         & 0.55        & 0.80         & 0.91        \\
\texttt{PC5}     & 0.59         & 0.49        & 0.90         & 0.82  \\
\hline
\end{tabular}
\end{table}

We also computed Popt for NPofB ranking, as reported in Table~\ref{tab:comp_popt_density}. There are still differences between the values for the two effort drivers, but they are not all in favor of McCC like in Table \ref{tab:comp_popt}.

\renewcommand{\arraystretch}{1}
\begin{table}[h]
	\centering
	\caption{Comparison of Popt values when LOC and McCC are used as effort drivers, when fault proneness is divided by size.}
	\label{tab:comp_popt_density}
		\begin{tabular}{l|cc}
			\hline
			& \multicolumn{2}{c}{Popt} \\
			Project & LOC & McCC  \\
			\hline
			\texttt{CM1}     & 0.68     & 0.72  \\
			\texttt{JM1}     & 0.72     & 0.69  \\
			\texttt{KC1}     & 0.73     & 0.73  \\
			\texttt{KC3}     & 0.68     & 0.66  \\
			\texttt{MC1}     & 0.57     & 0.58  \\
			\texttt{MC2}     & 0.80     & 0.78  \\
			\texttt{PC1}     & 0.69     & 0.79  \\
			\texttt{PC2}     & 0.63     & 0.69  \\
			\texttt{PC3}     & 0.67     & 0.69  \\
			\texttt{PC4}     & 0.75     & 0.82  \\
			\texttt{PC5}     & 0.86     & 0.80  \\
			\hline
		\end{tabular}
\end{table}
\renewcommand{\arraystretch}{1}

\subsection{Using Multiple Effort Drivers}\label{subsec:multiple}

Since the early research on software development effort~\cite{boehm1981software}, several studies have shown that effort depends on multiple factors, of which size and complexity are among the most relevant.
Accordingly, in computing an EAM, we could assume that the effort needed to analyze the $i^{th}$ module is
\begin{equation}\label{eq:compositeEffort}
\emph{effort}_i=\lambda\!\cdot\!s_i\!+\!(1\!-\!\lambda)\!\cdot\!q_i
\end{equation}
with $0\!\le\lambda\le\!1$.

\begin{figure}[h]
  \centering
  \includegraphics[scale=0.55]{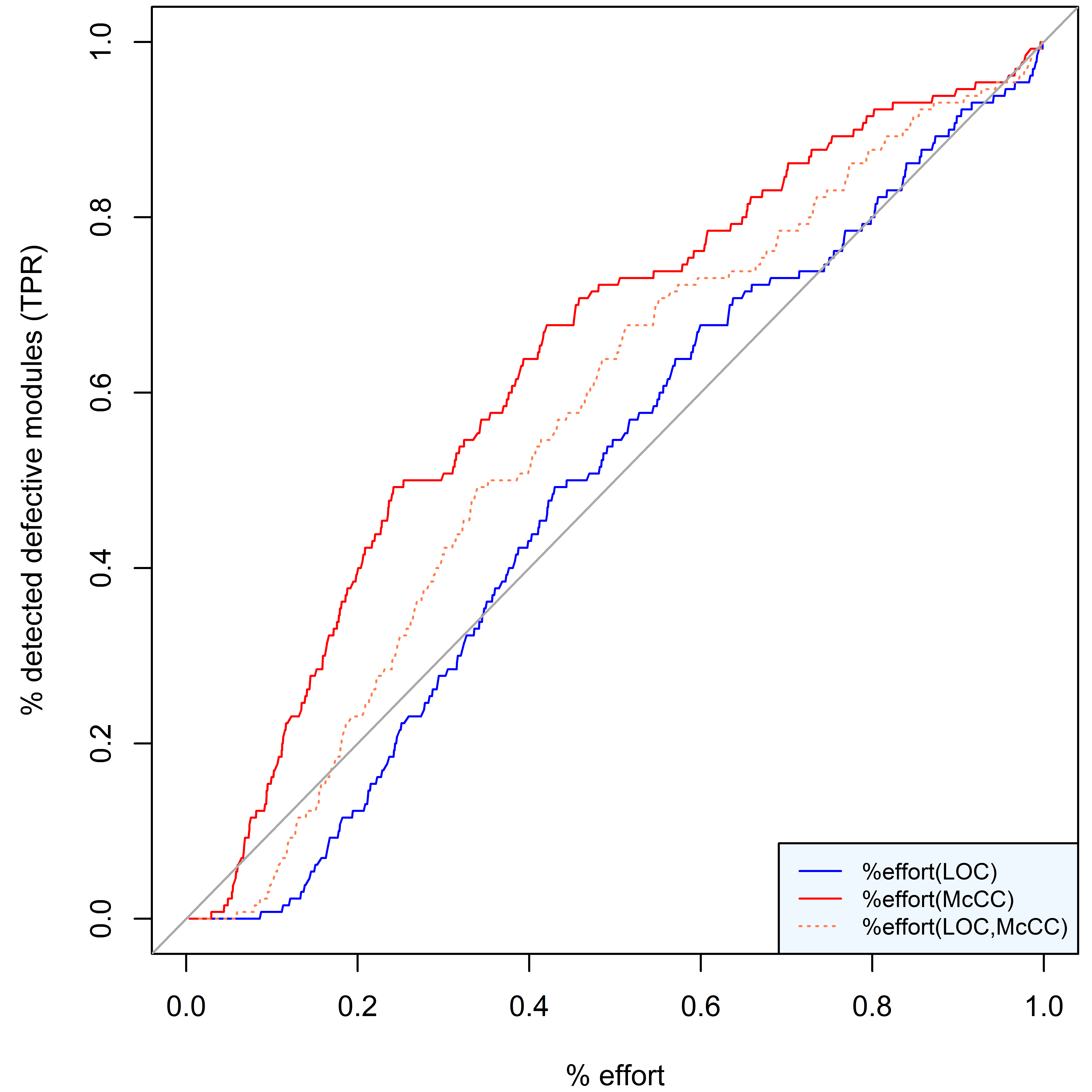}
  \caption{Cost efficiency curves of the BLR SDP model that uses LOC and McCabe's complexity density as independent variables for project \texttt{PC3}.}
  \label{fig:PC3COMPOSITE__LOC_TOTAL+CYCLOMATIC_DENSITY}
  \Description{}
\end{figure}

If we adopt this effort model with the size in LOC as $s$ and McCabe's complexity as $q$, we obtain cost efficiency curves that lie between those obtained assuming that effort depends on LOC and McCC, as shown in Figure~\ref{fig:PC3COMPOSITE__LOC_TOTAL+CYCLOMATIC_DENSITY}, where $\lambda=0.2$.
So, if effort actually depends on both size and complexity, assuming that it depends on just one of these code attributes leads to an optimistic and a pessimistic evaluations of the actual performance.

\subsection{Generalizing Results}

In this section, we consider a few characteristics of the study and we show that it is immediately generalizable as for  effort drivers, SDP models, and defect-count models.

\subsubsection{Using different effort drivers}
\begin{figure}[h]
  \centering
  \includegraphics[scale=0.55]{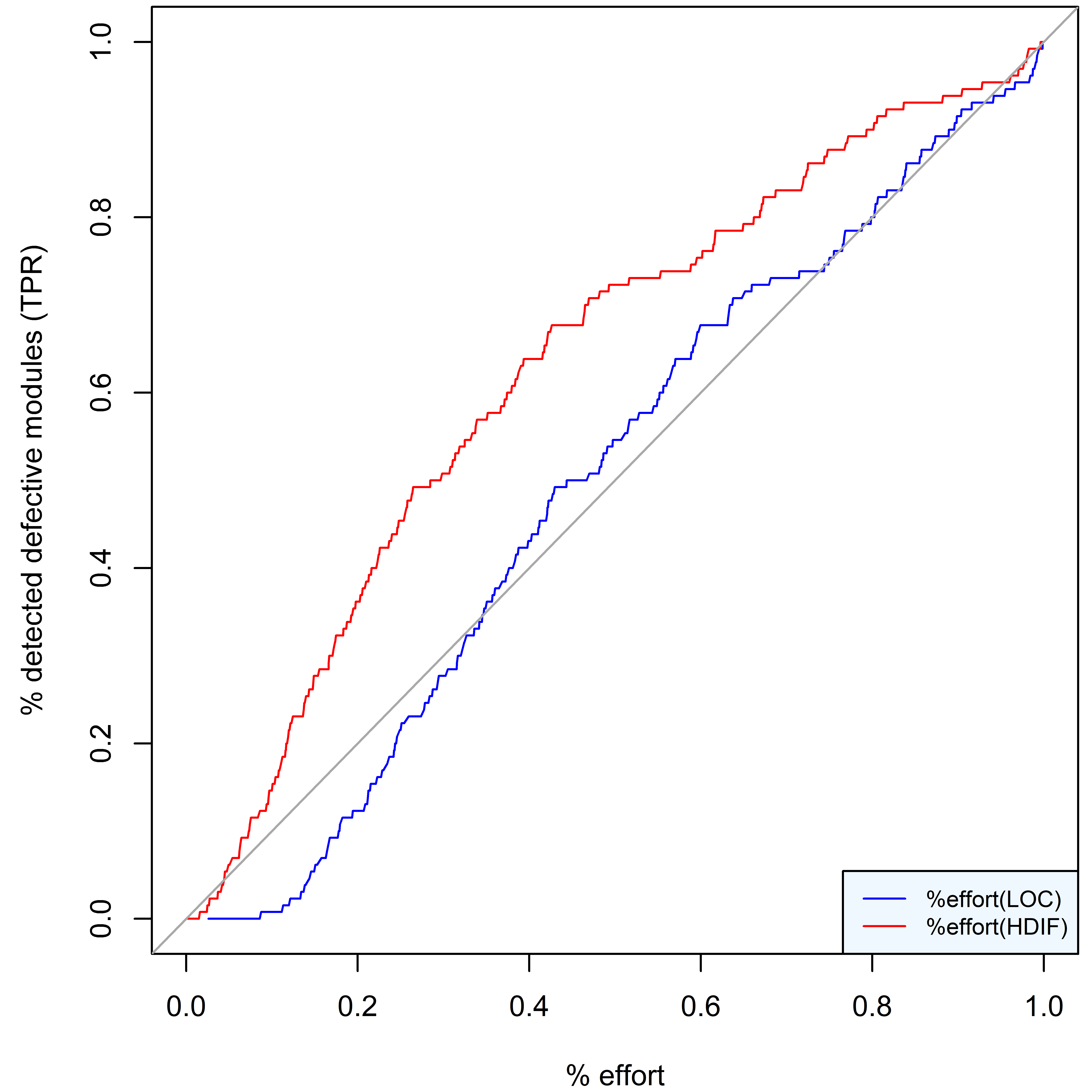}
  \caption{Cost efficiency curves of the BLR SDP model that uses LOC and McCabe's complexity density as independent variables for project \texttt{PC3}.}
  \label{fig:PC3_LOC_TOTAL+CYCLOMATIC_DENSITY_eff_HDIF}
  \Description{}
\end{figure}

In the sections above, we used McCabe's complexity as an alternative effort driver. What if module analysis effort is assumed proportional to a different code metric?
The argument given in Section~\ref{sec:otherMetrics} guarantees that cost efficiency curves are almost surely different. To get further evidence of the validity of the argument, we used Halstead Difficulty as an effort driver.
Halstead Difficulty purports to quantify the difficulty to write or understand the considered code, therefore
it is supposed to be correlated to module analysis effort.
Figure~\ref{fig:PC3_LOC_TOTAL+CYCLOMATIC_DENSITY_eff_HDIF} shows the cost efficiency curves of the same model used for Figure~\ref{fig:PC3_LOC_TOTAL+CYCLOMATIC_DENSITY}, but comparing LOC and Halstead Difficulty as effort drivers.
It is easy to see that assuming analysis effort proportional to Halstead Difficulty leads to quite different evaluations with respect to assuming effort proportional to LOC.

\subsubsection{Using a Different Fault-proneness Model}\label{subsubsec:differentModel}
\begin{figure}[h]
  \centering
  \includegraphics[scale=0.55]{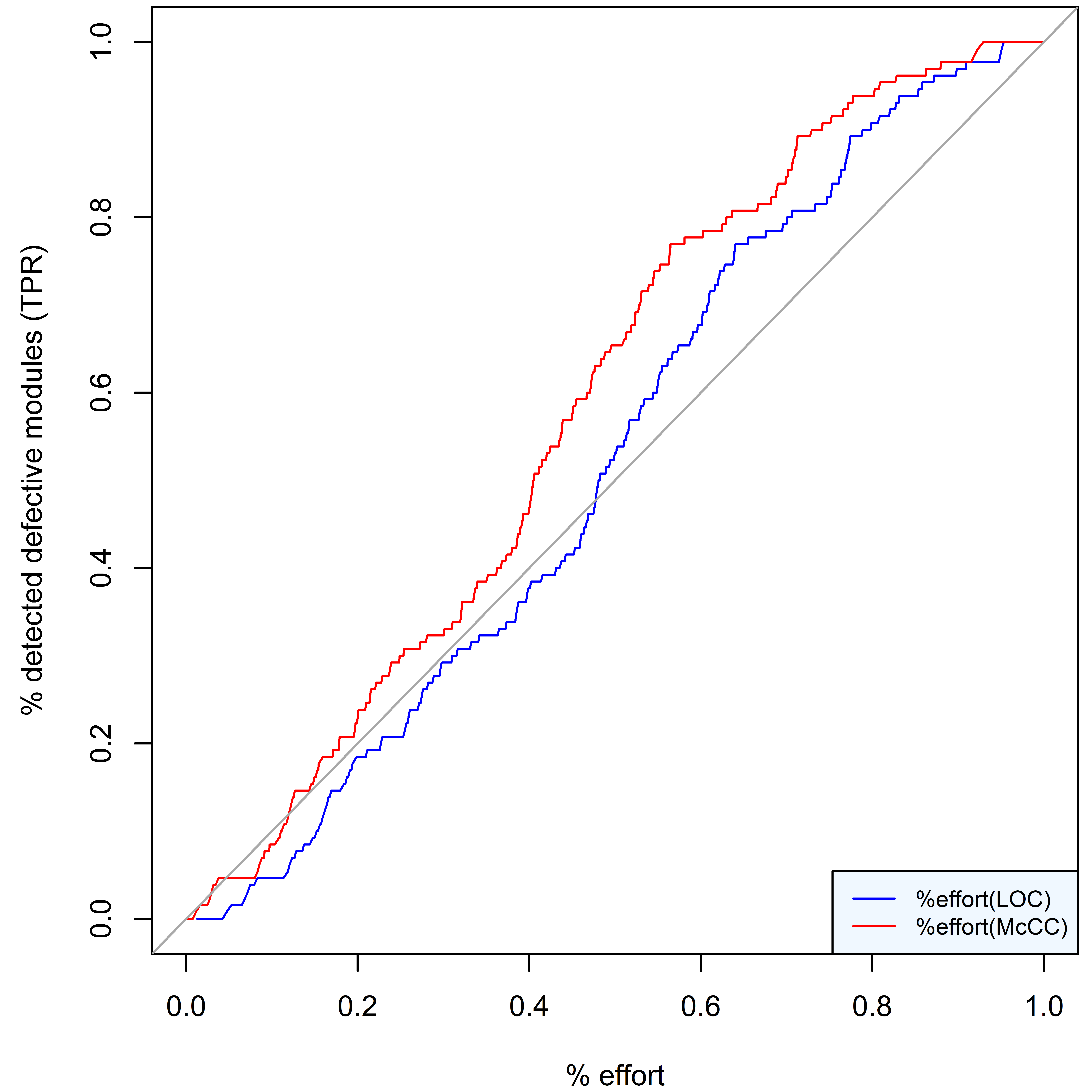}
  \caption{Cost efficiency curves of the BLR SDP model that uses Halstead vocabulary and number of operators as independent variables for project \texttt{PC3}.}
  \label{fig:PC3_HALSTEAD_VOCABULARY+NUM_OPERATORS}
  \Description{}
\end{figure}

One may also doubt that the presented results depend on the model used, especially because we used LOC and McCC density to build the fault-proneness model, and then used LOC and McCC as effort drivers.
Thus, we built a fault-proneness model for project \texttt{PC3} using Halstead Vocabulary and the Number of Operators as independent variables.
Figure~\ref{fig:PC3_HALSTEAD_VOCABULARY+NUM_OPERATORS} shows the cost efficiency curves for such model: the curves are clearly different when LOC and McCC are used as effort drivers.

\subsubsection{Evaluating Models That Estimate the Number of Defects per Module}\label{subsubsec:defectnumest}
In the sections above we used BLR to build defectiveness probability models by analyzing datasets that report whether each module contains at least one defect.
When the available dataset contains the number of defects of each module, it is possible to build models that predict the number defects per module. So, one could wonder whether the dependency of EAMs from effort models applies also to models that predict the number defects per module.

To this end, we can observe that the available effort determines the number $EP$ of estimated positive modules, hence the performance, given a module ranking. That is, $EP$ depends on the available effort.
The argument discussed in Section~\ref{sec:otherMetrics} applies to any module ranking, hence we expect that EAMs for models that predict the number defects per module yield different indications depending on the code measure used as an effort driver.

To get a practical demonstration of this, we built Random Forest (RF) regression models using some of the datasets from the collection by Jureczko and Madeyski~\cite{jureczko2010towards}. The obtained models predict the number of defects per module.
The modules were then ranked by decreasing number of predicted defects; modules having the same number of predicted defects were ordered by decreasing LOC. Cost efficiency curves were then built: they yield the proportion of defects found as a function of the percentage of effort allocated to module analysis.

\begin{figure}[h]
  \centering
  \includegraphics[scale=0.55]{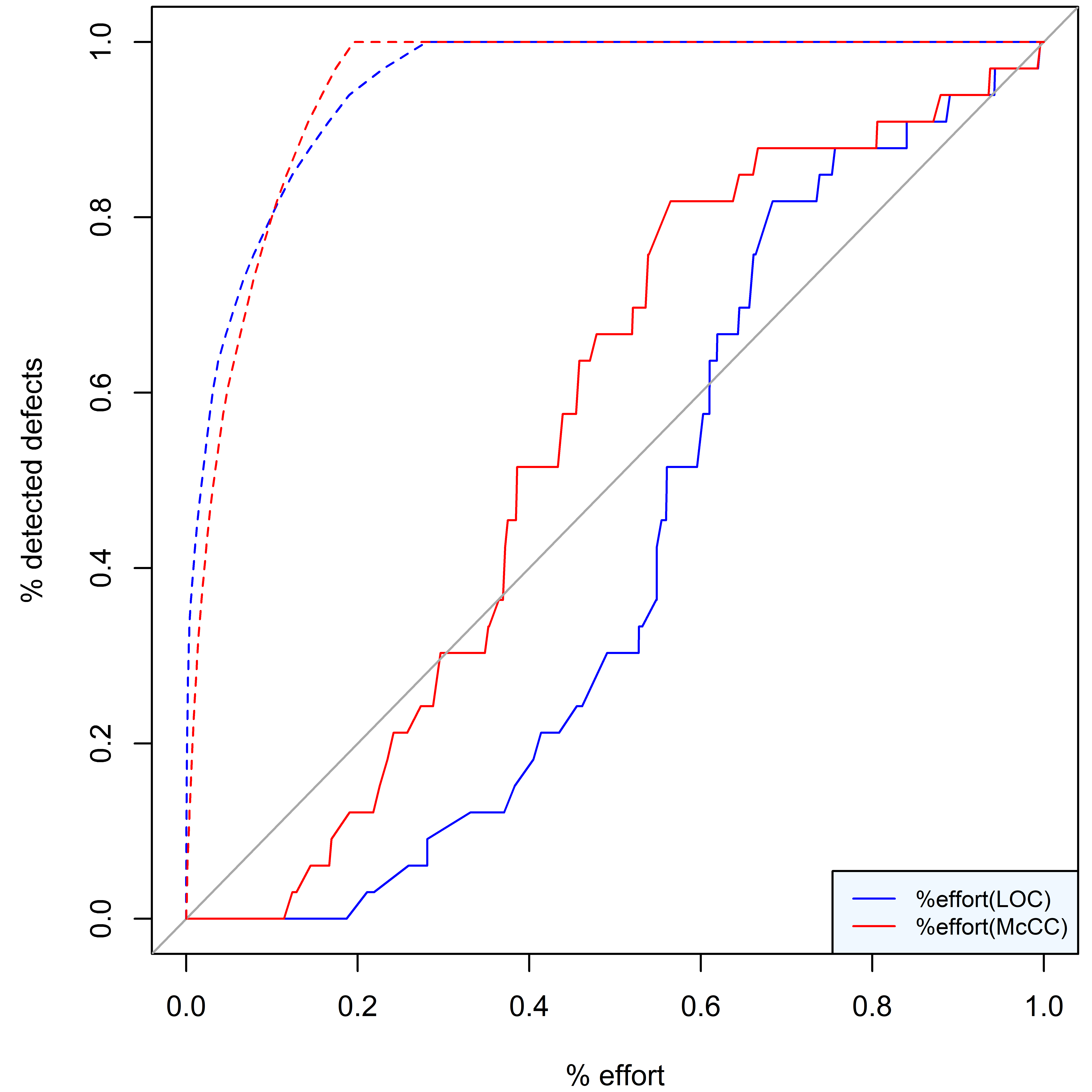}
  \caption{Cost efficiency curves of RF regression model that uses LOC and McCC density as independent variables and the optimal model for project \texttt{arc}.}
  \label{fig:arc_loc_McCC_density_ideal}
  \Description{}
\end{figure}

Figure~\ref{fig:arc_loc_McCC_density_ideal} shows the cost efficiency curves of the obtained RF regression model and the optimal model for project \texttt{arc}. It is quite clear that the curves are different when LOC and McCC are used as effort drivers.

\section{Threats to Validity}\label{sec:threats}
Since the results described in the paper descend from the properties discussed in Section~\ref{sec:otherMetrics}, the results of the empirical study are hardly subject to relevant threats, as discussed below.

\textit{External Validity.} Different models and different datasets would result in new rankings that could yield different EAM values. However, as discussed in Section~\ref{sec:otherMetrics}, the cost efficiency curves generated using different code measures as proxy for effort would be different, in general, regardless of the model and dataset.
Different models and datasets can only determine to what extent cost efficiency curves associated with different effort models differ.

As mentioned in Section~\ref{subsubsec:dataset}, the datasets from the NASA collection we used do not provide the number of defects in each module. Thus, our cost efficiency curves and the computations of PofB, NPofB and Popt refer to the proportion of faulty modules detected, rather than the proportion of faults detected.
Nonetheless, both the argument in Section~\ref{sec:otherMetrics} and the results described in Section~\ref{subsubsec:defectnumest} show that EAMs depend on the effort model also when dealing with the proportion of faults detected.

\textit{Internal Validity.} The way the models were implemented can threaten internal validity. We built our Logistic Regression models in {\texttt{R} using the \texttt{glm} function.
This code is generally considered very reliable, hence we do not see any real threat concerning the internal validity of our study.

\section{Related Work}\label{sec:related}
Arisholm et al. argued that module analysis effort is proportional to size. Specifically, they argued that inspection cost is proportional to size, but, depending on the type of inspection, one should use a different type of size measure, like the number of functions, lines of code, and also control flow complexity~\cite{arisholm2006predicting}. In a second paper they argued that the effort to inspect and test a class is proportional to size, in terms of Number of Statements~\cite{arisholm2007data}.

Mende and Koschke suggested that effort can be better correlated with other code measures than size: they used McCabe's complexity as a proxy for effort in their seminal paper~\cite{mende2010effort}. Nonetheless, some papers (e.g.,~\cite{kamei2010revisiting}) erroneously report that Mende and Koschke used lines of code as a proxy for effort.
As a result, the paper by Mende and Koschke is very often cited to justify the assumption that module analysis effort is proportional to the size in LOC.

PofB and Popt are by far the two most used effort-aware metrics~\cite{ccarka2022effort}. There are, however, other EAMs that are used in literature. Huang et al. proposed IFA (Initial False Alarms): the number of initial false alarms encountered before the first real defective module is found~\cite{huang2017supervised, huang2019revisiting}.
Huang et al. also proposed PCI@20\%, which represents the proportion of inspected changes when 20\% of LOC are inspected. This metric is meant to be used in a Just-In-Time context, but it can also be applied in other contexts. In these cases, it is usually referred to as either PMI or PFI, which are the proportion of modules and files inspected, respectively~\cite{chen2021revisiting,qu2021leveraging}.
These three metrics are meant to represent the amount of context switches that developers have to perform while inspecting the code, with lower values of PCI/PMI/PFI being preferable.

Researchers have also measured the performance of models by using traditional performance metrics in an effort-aware fashion, when only a percentage of LOC is inspected. E.g., Precision@20\% and Fscore@20\% are the Precision and F-score achieved when 20\% of the code LOC have been analyzed~\cite{huang2017supervised,xu2018cross}.
Bludau and Pretschner have also argued in favor of the usage of negative performance metrics, namely Negative Precision and True Negative Rate~\cite{bludau2024towards}.

The ranking of modules is crucial in determining the performance of SDP models. There are other ways to optimize models based on different effort-aware performance metrics. For example, Chen et al. proposed a logistic regression model whose coefficients are optimized using a multi-objective optimization algorithm (NSGA-II)~\cite{chen2018multi}.
However, EAMs are bound to yield different indications with different effort models, regardless of the ranking criteria, as we showed above (see Sections~\ref{subsec:npofb} and~\ref{subsubsec:defectnumest}).




\section{Conclusion}\label{sec:conclusion}
Effort-aware metrics (EAMs) are generally based on the assumption that effort is proportional to size, measured in LOC.
Some papers have warned against this assumption, based on the observation that the effort required to analyze modules for finding defects is often better correlated to other code metrics, also depending on how module analysis is carried out~\cite{mende2010effort,shihab2013lines}.

In this paper, we study the behavior of the most popular EAMs---namely PofB (together with its normalized version NPofB) and Popt---when analysis effort is assumed to be proportional to different code metrics.

We show that, given a module ranking,
assuming that effort depends on different code measures is bound to yield different indications about performance, in general.
Even though the argument we provide is sufficient to explain the different evaluations that result from assuming different effort drivers, we carried out an empirical study that illustrates the practical consequences of the dependence of EAMs on the underlying effort model.
Specifically, we showed that different assumptions regarding effort drivers result in i) different optimal cost efficiency curves, ii) different cost efficiency curves for a given model and ranking, iii) different values of PofB and NPofb, for any given threshold value, iv) different values of Popt.
%

Researchers and practitioners should be informed that i) the evaluations yielded by EAMs depend on the underlying effort model, and different effort models lead to conflicting conclusions; ii) the EAMs used until now are actually LOC-aware, rather than effort-aware, hence they cannot be trusted in general: e.g., there are many module analysis activities that require an amount of effort better correlated to McCabe's complexity than to LOC.

To correctly apply EAMs, we should use an accurate effort model that accounts for all the factors that affect effort, otherwise excessively optimistic or pessimistic performance evaluations are obtained.
Interestingly, the concepts underlying EAMs do not depend necessarily on an effort model: it is sufficient to associate each module with a reliable evaluation of its analysis effort---instead of its size in LOC---to get measures that are really effort-aware. Actually, the better the effort model, the more reliable the evaluation provided by EAMs.

\section*{Supplemental Material}
More results of our empirical studies are at \url{https://github.com/BruhZul/Supplement-material-EAM-EASE25}.


\bibliographystyle{ACM-Reference-Format}
\bibliography{theBib}

\end{document}